# Fabrication of phononic filter structures for far-IR/sub-mm detector applications

Kevin L. Denis*[a], Karwan Rostem[a], Marco A. Sagliocca[a], Elissa H. Williams[b], Edward J. Wollack[a]
[a]NASA Goddard Space Flight Center, 8800 Greenbelt Road, Greenbelt, MD 20771;
[b]Science Systems and Applications, 10210 Greenbelt Road, Lanham, MD 20706

## ABSTRACT

A photon noise limited sub-mm/far-IR cold telescope in space will require detectors with noise equivalent power (NEP) less than 1x10$^{-19}$ W/Hz$^{1/2}$ for imaging applications and at least an order of magnitude lower for spectroscopic studies. The detector NEP can be reduced by lowering the operation temperature and improving the thermal isolation between the bolometer and a heat bath. We report on the fabrication of membrane isolated transition edge sensor bolometers incorporating compact (<50 µm) thermal isolation beams based on phononic filters. Phononic filters are created by etching quasi-periodic nanoscale structures into supporting thermo-mechanical beams. The cross-sectional dimensions of the etched features are less than the thermal wavelength at the operating temperature, enabling coherent phonon transport to take place in one dimension. The phonon stop-band can be tuned by adjusting the scale of the quasi-periodic structures. Cascading multiple filter stages can increase bandwidth and provide improved thermal isolation similar to the function of a multi-stage electrical filter. We describe the fabrication of AlMn based transition edge sensor bolometers on silicon and silicon nitride membranes isolated by one- and two-dimensional phononic filters. The phononic filters are patterned through electron beam lithography and isolated with deep reactive ion etching.

## 1. INTRODUCTION

Superconducting transition edge sensor bolometers have demonstrated background limited performance for ground based and sub-orbital applications in astronomy. The next generation of far-infrared space-borne telescopes will require noise equivalent power <1x10$^{-19}$ W/Hz$^{1/2}$ for imaging applications and ~1x10$^{-21}$ W/Hz$^{1/2}$ for spectrometry [1]. For a transition edge sensor (TES) bolometer, the noise equivalent power (NEP) is given by NEP = $\sqrt{4k_B T_c^2 G_{th}}$, where $T_c$ is the superconducting TES transition, and $G_{th}$ is the thermal conductivity between the sensor and the thermal bath. The NEP is commonly reduced through reduction of the thermal operating point $T_c$. For space applications, typical values on the order of 0.1 K are limited by the capabilities of adiabatic demagnetization refrigerator systems that provide a bath temperature of approximately 50 mK for detector arrays [2]. Further reduction in NEP can be achieved through reduction in the thermal conductance between the sensor and the bath. The simplest conceptual approach is a bolometer with long and narrow support beams for isolation. NEPs on the order of 1x10$^{-19}$ W/Hz$^{1/2}$ have been achieved in this way [3,4,5]. However, drawbacks of this approach include the reduced filling fraction of the sensor's absorber structures and the potential for excess heat capacity distributed along the beams. Additionally, repeatable thermal response may be limited by phonon scattering off the potentially variable and difficult to control roughness along the beams [6]. Low thermal conductivity has also been achieved through the use of hot electron effects in nano-bolometers [7] with reduced electron-phonon coupling achieving a NEP approaching 1x10$^{-19}$ W/Hz$^{1/2}$. Here we utilize the fact that phonons can propagate coherently through a solid medium at temperatures below ~200 mK. It is therefore possible to design structures such that there is a phononic bandgap that blocks modes of heat flow through the beams [8,9,10]. A previous paper describes our design approach, which consists of incorporating multistage filters with tuned bandgaps that can be thought of as being analogous to poles in an electronic filter. A stop-band has been designed to provide high rejection of heat flow in a compact geometry [11]. A variety of phononic device geometries have been fabricated including 2D meshes as well as 1D structures articulated by bends and T-shaped features. TES devices were fabricated on a beam-isolated 110 µm x 140 µm membrane. In this paper, we report on the fabrication process used to realize these structures and provide a brief status of device characterization.

*kevin.l.denis@nasa.gov; phone 1-301-286-7935; fax 301-286-1672; nasa.gov

## 2. FABRICATION

### 2.1 Fabrication process

The simplified fabrication process is shown in Figure 1. Table 1 lists the material layers and process details. The process starts with a silicon wafer coated with silicon dioxide and low stress silicon nitride. The silicon nitride is used as the membrane and structural support for the detector. The silicon nitride is grown by low-pressure chemical vapor deposition as low stress (< 150MPa) and slightly tensile. A tensile stress is desirable for quarter-wave backshort absorber-coupled detectors where a membrane in tension leads to higher pixel flatness and improved optical coupling efficiency[12]. The first step in the fabrication process is lift-off of the contact and electron beam lithography alignment marks. We use AZ-5214E image reversal photoresist and deposit a thin layer of titanium (5 nm) and gold (100 nm). Gold is used for high contrast in the electron beam lithography system. The next step is to define niobium leads and the gold contact pads. This step is done as a lift-off process, again using 5214E image reversal. The Nb (50 nm) and Au (10 nm) are deposited by DC sputter deposition sequentially to improve the electrical contact between the layers. The sputter deposition of the niobium and gold was optimized to ensure a low stress of less than 100 MPa to reduce potential bowing of the membrane and beams. Next, the wafer is patterned with positive S-1811 photoresist, which protects the gold in the areas of wire-bond pads and where the niobium makes contact with the TES. The gold is used to ensure that the niobium does not oxidize during the subsequent processing steps. Next, the wafer is patterned for electron beam lithography of the phononic structures. We use ZEP 520A electron beam photoresist spun to a thickness of 450 nm. The wafers are exposed at 100kV in a JEOL JBX 6300-FS direct write electron beam lithography system at the National Institute of Standards and Technology Center for Nanoscale Science and Technology (NIST CNST, Gaithersburg, MD). After patterning, the wafers are etched in a capacitive coupled plasma reactive ion etching tool in a $CHF_3:SF_6$ (3:1) plasma chemistry at 100 W and 20 mT based on the recipe in [13]. The low operating pressure increases the plasma self-bias increasing etch directionality. Further, the addition of fluorocarbon polymer passivates the feature sidewalls to reduce lateral etching and increase anisotropy. Approximately a 1:1 etch selectivity to the resist is achieved. After etching, the photoresist is ashed in an $O_2$ plasma and cleaned in Plasma-Solve EKC solvent to remove residual fluorocarbons from the resist and process chemistry. The etch process terminates on the silicon dioxide layer. A similar process can be used with single crystal silicon via silicon-on-insulator wafers.

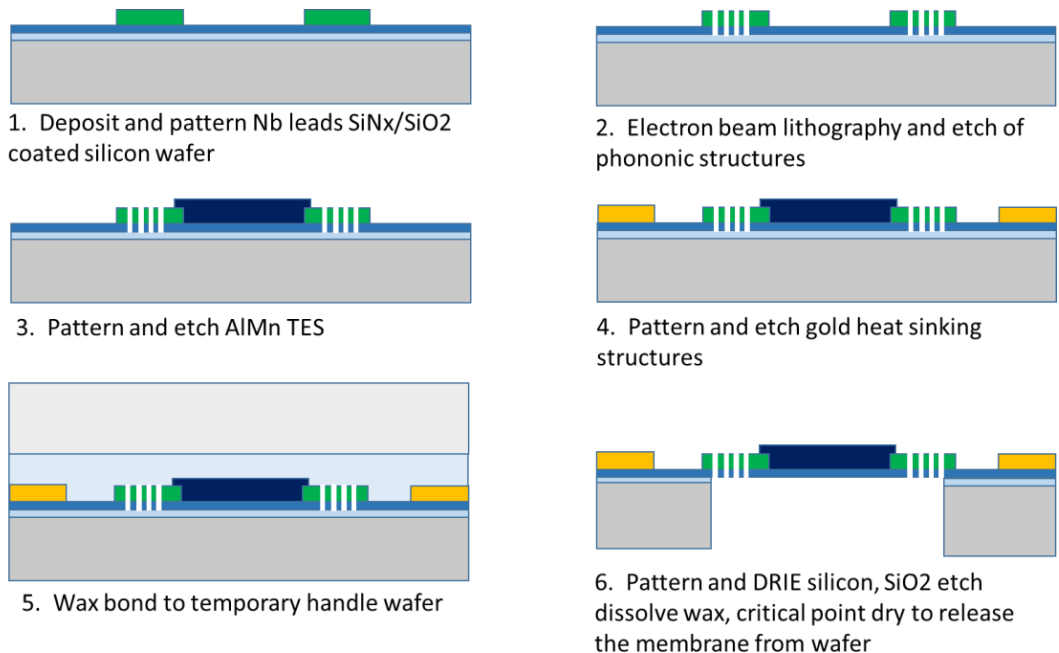

Figure 1. Fabrication procedure.

Minimum phononic structure feature sizes on the order of 75 nm are achieved with this process. Depending on the geometry, final structures are typically offset from the mask by 15 – 20 nm, indicating minimal resist undercut. The gold film for chip heat sinking and TES detectors is patterned with AZ-5214E and electron beam deposited for lift-off. Next, we deposit the AlMn TES via a lift-off patterning step using 5214E on those pixels that are TES based. The niobium leads are pre-cleaned with an in-situ reverse bias RF sputter to remove $NbO_x$ then a 200 nm thick AlMn film is DC sputtered from an alloy target with 3000 ppm atomic percent. This is the required percent Mn to give a transition temperature of approximately 75 mK [14]. The film lift-off is performed in acetone. Next, the street cut is defined by contact-based photolithography and the silicon nitride is etched in $CHF_3/SF_6$ plasma completing front side processing. For our first fabrication run, the street cut defined both the outside edge of each chip, singulated in a backside DRIE (Deep Reactive Ion Etch) process, as well as the membrane etch for the pixels. Both the phononic patterns and the membrane definition could have been done by electron beam lithography. To reduce the e-beam writing area and time, these steps were achieved separately. The electron beam lithography was done in only the areas of the support beams, and areas connecting the beams to the membrane were etched using contact lithography. The drawback of this process is that the two etches had overlap to account for any misalignments. This created a small area that was double etched leaving a pit in the silicon oxide etch stop. The wafer was then bonded to a Pyrex handle wafer with wax. It was found during subsequent processing that the double etched pit in the $SiO_2$ could trap voids in the wax and might cause additional bending and breakage of the membrane. To avoid this problem, future wafers will use electron beam lithography to fully define each pixel. After wax bonding, the backside of the silicon is patterned with thick SPR220-7 photoresist. The 525 µm thick silicon wafer is etched down to the buried oxide with a standard Bosch process consisting of alternating $SF_6$ and $C_4F_8$ cycles. The silicon dioxide functions as an etch stop which is removed in HF solution. Finally, the wax is dissolved in acetone and the chips are separated from the Pyrex wafer and dried in a $CO_2$ critical point dryer.

| Material | Thickness [nm] | Process | |
|---|---|---|---|
| Niobium | 50 | DC Sputter | Etch: $SF_6/CHF_3$ at 100 W, 20 mT |
| Gold contact | 15 | DC Sputter | KI etch |
| Gold heat sink / heater | 300 | Electron beam evaporation | Lift-off AZ-5214E in acetone |
| AlMn TES | 400 | DC Sputter | Lift-off AZ-5214E in acetone, target 70 mK transition temperature |
| $SiN_x$ membrane | 100 | LPCVD | Etch: $SF_6/CHF_3$ at 100 W, 20 mT |
| $SiO_2$ etch stop | 300 | Thermal Oxidation | Buffered HF (7:1) |
| Silicon | 525 (um) | - | BOSCH $SF_6$, $C_4F_8$ |

Table 1. Material thicknesses and process information. Note that a similar process could be achieved using ultra-thin silicon-on-insulator wafers where the $SiN_x$ membrane is replaced by single crystal silicon.

## 3. RESULTS

### 3.1 Yielded Pixel Designs

A variety of pixel designs were fabricated to validate the fabrication feasibility, device yield, and device performance across a broad design space. The designs are shown in Figure 2. At the operating temperature of the TES, the phonon wavelength ($\gtrsim 1$ µm) is larger than the thickness and width of the largest unit cell in each design. This ensures that phonon propagation is along the direction of the beam (1D). Early devices were bonded directly to a Pyrex wafer with wax and released at the wafer level as described above. The release process consisted of dissolving wax in acetone. The separated chips were cleaned in acetone and isopropyl alcohol then dried on a hotplate or by critical point drying. We found that this process resulted in a very low yield due to broken membranes. To improve this process, subsequent wafers were bonded to a Pyrex wafer with 5 µm tall SU8 posts designed to align to the field region of each chip. These wafers were diced after the DRIE etching step and the release step was completed individually at the chip level. The SU8 posts served two functions. First, they set the wax thickness to a repeatable value and second, they provided an offset that keeps the released membranes from touching the Pyrex during the releasing process. Consequently, this allowed release through the critical point drying step to be completed without removing the Pyrex handle wafer from the device wafer. The previous process required the Pyrex wafer be removed from the device wafer prior to the cleaning step by sliding the device chip off of the Pyrex wafer while in the acetone solution. This additional handling step/sliding process damaged the released membranes. In the new process, after critical point drying, the Pyrex would sit on top of the SU8 posts without making contact with the released membranes and could simply be lifted off of the released chip. Wax corrugations caused by plasma heating were observed after DRIE. This is consistent with previous fabrication processes [15] where it was determined that the temperature of the DRIE process can cause reflow and introduce ripples in the wax, which bend the membranes and break the support structures.

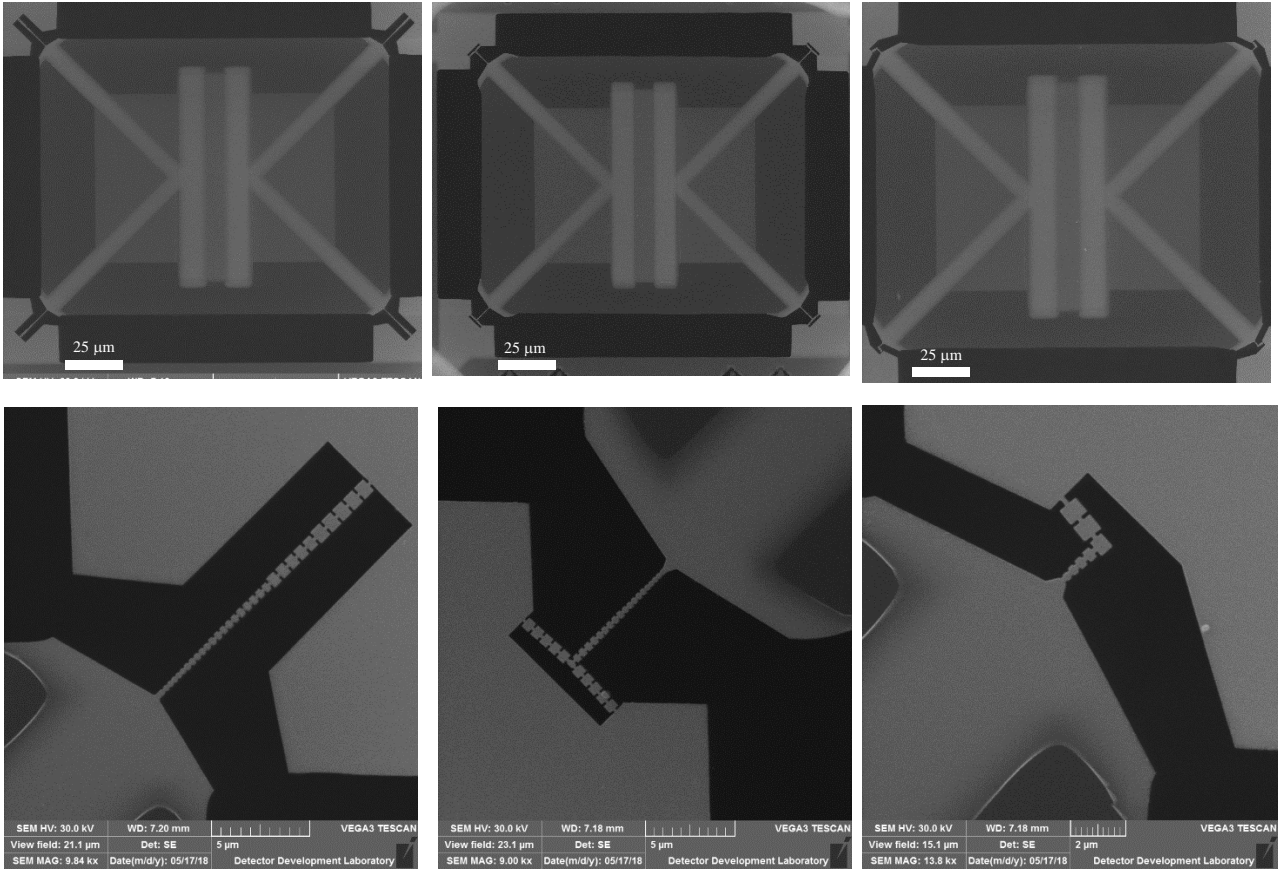

Figure 2. Scanning electron micrographs of pixel designs that completed full fabrication process. It was found that membrane bowing was the likely cause of yield loss. The structures shown above were flexible enough to accommodate slight bowing of the TES membrane. Minimum feature sizes on these structures are 75 nm.

Additionally, it was found that the $SiO_2$ HF etching process could etch some oxide membranes faster than others on the same chip. This was likely from insufficient wetting along the sidewalls of the backside high aspect ratio silicon etch and resulted in air bubbles being trapped in some structures. The over-etched structures in turn resulted in front side membrane niobium being attacked due to undercut of the wax. This problem was worse when an air bubble was trapped in the wax around the beam due to the topography of the wafer. Simply using a surfactant prior to HF etching solved this problem on subsequent chips.

Of the designs shown in Figure 2, the 15 μm long straight structures and the 8 μm long "Tee"-shaped structures had the highest yield. In fact, the straight structure device yielded even when chips were simply allowed to dry in air on a hotplate rather than using the critical point dry. The most compact structure to complete fabrication is shown to the right in Figure

2, where a bend in the beam serves as a mode conversion junction that scatters out-of-plane flexural modes into torsional modes and compression into in-plane flexural modes [11]. While careful handling and releasing methods described here significantly increased yield, the very short/rigid thermal isolation structures still tended to break. We suspect that the low yield for these structures is caused by film stress bending the membranes out of plane by approximately 3 µm. The resulting strain breaks the shortest beams while the longer beams are flexible enough to compensate. In Figure 2, the slight bending in the Tee-structure and the bent beam can be seen due to the curvature of the membrane.

Further confirmation of this hypothesis is shown by an additional structure that was built by first completing the fabrication process but without patterning and etching the phononic structures. The wider free-standing support structure remained and could then be etched directly by focused ion beam. Etching of phononic structures by focused ion beam was performed at NIST CNST, Gaithersburg, MD and the resulting device is shown in Figure 3 below. It is demonstrated that the smaller section phononic structures, < 5 µm long, which are compatible with e-beam lithography, can be yielded if embedded in a compliant support structure on the order of 20 µm long. To improve the filling fraction, future devices could include a short meandered section with an embedded phononic filter.

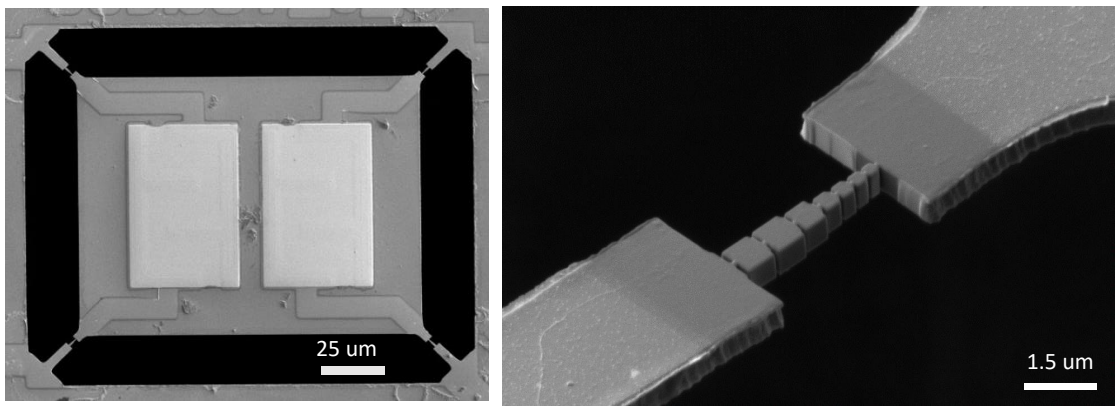

Figure 3. Scanning electron micrograph of a focused ion beam fabricated single Johnson noise-based test pixel and magnified image of one of the phononic thermal isolating structures etched into the free-standing beams. Residual gallium contamination can be seen as the dark area coating about 2 µm of the beam.

## 4. TESTING

Progress has been made in building an ultra-low-noise package for testing of the phononic-isolated devices and TES detectors. A custom chip-scale 3-pole RC filter was procured from Mini-Circuits Included (MSI [16]). The RC filters have a 20 kHz 3 dB cutoff frequency. An absorptive filter with an attenuation constant of 16 dB/GHz ensures high frequency radiation is severely attenuated before reaching the device under test [17]. The absorptive filter is stainless steel powder that is bonded to the package using cyanoacrylate (superglue). The copper package, absorptive filter, and custom made silicon fanout boards have been fabricated. Testing will be completed in a BluFors cryogen free dilution refrigerator with a base temperature of less than 10 mK providing greater than 250 µW of cooling at 100 mK. The test package and filter response is shown in Figure 4.

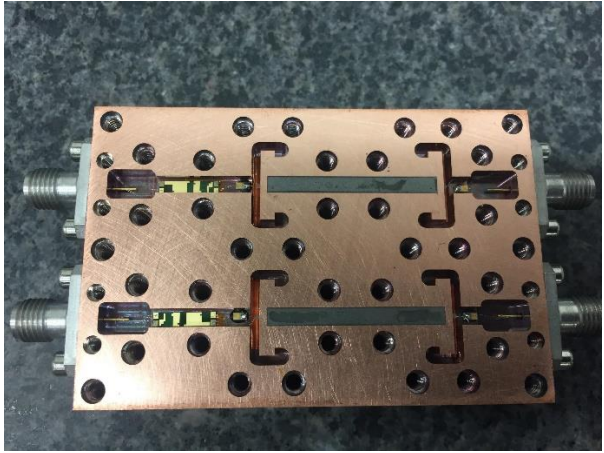 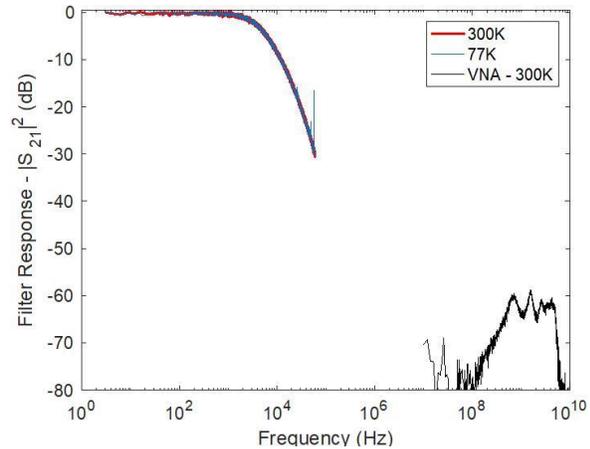

Figure 4. Photo of the low noise package consisting of chip based RC filters [16] and powder based absorptive filter [17]. Also shown is room temperature and cryogenic response of the package.

## 5. CONCLUSION

A fabrication process was described to integrate transition edge sensors into a bolometer structure isolated by phononic filters designed for low thermal conductivity. The process is based on our standard detector fabrication processes demonstrated for HAWC+, CLASS, BETTII, HIRMES, and ATHENA with the incorporation of an electron beam lithography step to realize nanoscale phononic structures. A variety of structures have been demonstrated including 2D, 1D, and 90 degree bent structures, which are yielded through careful control of the membrane release and handling processes. Short 1D structures can be embedded in a longer compliant beam patterned by standard contact lithography to improve the fabrication yield due to membrane bowing. Our current devices have minimum feature sizes of 75 nm with ~ 15 nm of undercut using a high aspect ratio etching process incorporating $CHF_3$ and $SF_6$ based chemistry. The process can also be utilized to build similar structures using silicon-on-insulator wafers where the $SiN_x$ membrane is replaced with a silicon membrane. The electron beam lithography writing time for our test wafers, which consisted of 256 pixels, required approximately 30 minutes. For a flight-format detector array on the order of 10,000 pixels, total electron beam lithography writing times are estimated at 19 hours per array. This is reasonable given the low throughput required for custom detector fabrication. Future devices will incorporate SOI wafers. A cryogenic deep reactive ion etching system has been installed in the Goddard Space Flight Center Detector Development Laboratory, which will lead to improved etching capabilities for the nanoscale etching of silicon-based structures down to 50 nm and below. A low noise cryogenic testing package has been described. Cryogenic thermal testing is currently underway.

## ACKNOWLEDGEMENTS

This work was supported by the NASA Astrophysics Research and Analysis (NNX17AH83G) and Goddard Space Flight Center Internal Research and Development Programs. We gratefully acknowledge Richard Kasica and Joshua Schumacher at NIST Gaithersburg Center for Nanoscale Science and Technology for their support on electron beam lithography and focused ion beam milling.